\documentclass{PoS}

\usepackage{url}
\def\arxiv#1{\href{http://arxiv.org/abs/#1}{{\tt arXiv:#1}}}

\title{H.E.S.S. observations following multi-messenger alerts in real-time}

\ShortTitle{H.E.S.S. multi-messenger observations}

\author{\speaker{Fabian Schüssler}\\
        IRFU, CEA, Université Paris-Saclay, F-91191 Gif-sur-Yvette, France\\
        E-mail: \email{fabian.schussler@cea.fr}}
\author{ M.~Backes, A.~Balzer, F.~Brun, M.~F\"u{\ss}ling, C.~Hoischen, J.-P.~Lenain, I.~Lypova, S.~Ohm, D.~Parsons, G.~P\"uhlhofer, A.~Reimer, G.~Rowell, M.~Seglar-Arroyo, A.M.~Taylor on behalf of the H.E.S.S. collaboration\\ E-mail: \email{contact.hess@hess-experiment.eu}}    

\abstract{The H.E.S.S. Imaging Air Cherenkov Telescope system is, due to its fast reaction time and its comparably low energy threshold, very well suited to perform follow-up observations of detections at other wavelengths or other messengers like high-energy neutrinos and gravitational waves. These advantages are utilized optimally via a fully automatized system reacting to alerts from various partner observatories covering various wavelengths and astrophysical messengers.
In this contribution we'll provide an overview and present recent results from H.E.S.S. programs to follow up on multi-wavelength and multi-messenger alerts. To illustrate the capabilities of the system we present several real-time ToO observations searching for high-energy gamma-ray emission in coincidence with high-energy neutrinos detected by the IceCube and ANTARES neutrino telescopes and outline our program to search for gravitational wave counterparts.}

\FullConference{35th International Cosmic Ray Conference --- ICRC2017\\
		10--20 July, 2017\\
		Bexco, Busan, Korea}

\begin{document}
\section{High-energy multi-messenger astrophysics}
After decades of ever intensifying studies we are still searching for the origin of the bulk of high-energy cosmic rays. On the other hand significant progress has been made over the last years: thanks to observatories of unprecedented scale and sensitivity a wealth of discoveries has been provided by gamma-ray observatories sensitive in the GeV and TeV energy range, the advent of neutrino astronomy  was brought about by large-scale neutrino telescopes and the direct observation of gravitational waves completed the multi-messenger picture of the high-energy universe.

In this context, the main focus of the H.E.S.S. multi-messenger program is to exploit the intimate connection between these new messengers and gamma-rays. In the fortunate case that the environment of the cosmic accelerators exhibits appropriate conditions, high-energy (hadronic) particles are potentially undergoing interactions with matter and radiation fields within and/or surrounding the acceleration sites. The interaction products (mainly light mesons like pions) will decay by emitting both high-energy neutrinos as well as gamma-rays. Similar mechanisms are expected to play a role in gravitational wave emitting mergers of compact objects like binary neutron stars. For local sources for which the matter and radiation fields are not too dense to cause absorption of the emitted gamma-rays, we can hope to find spatial and temporal correlated emission of different messengers. 


\begin{figure}[!t]
\vspace{-5mm}
\begin{minipage}{0.375\linewidth}
\centerline{\includegraphics[width=0.95\linewidth]{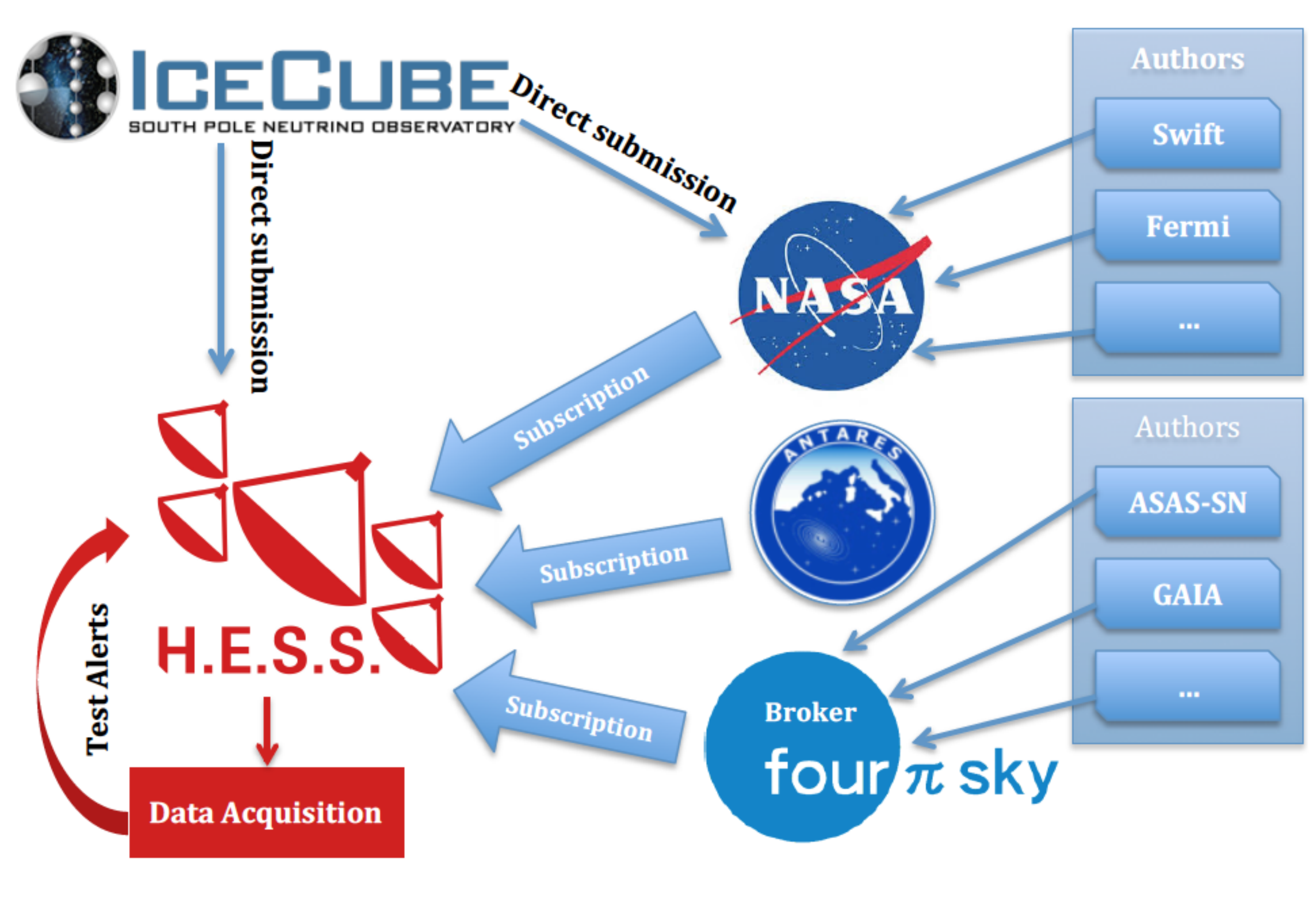}}
\end{minipage}
\hfill
\begin{minipage}{0.625\linewidth}
\centerline{\includegraphics[width=0.95\linewidth]{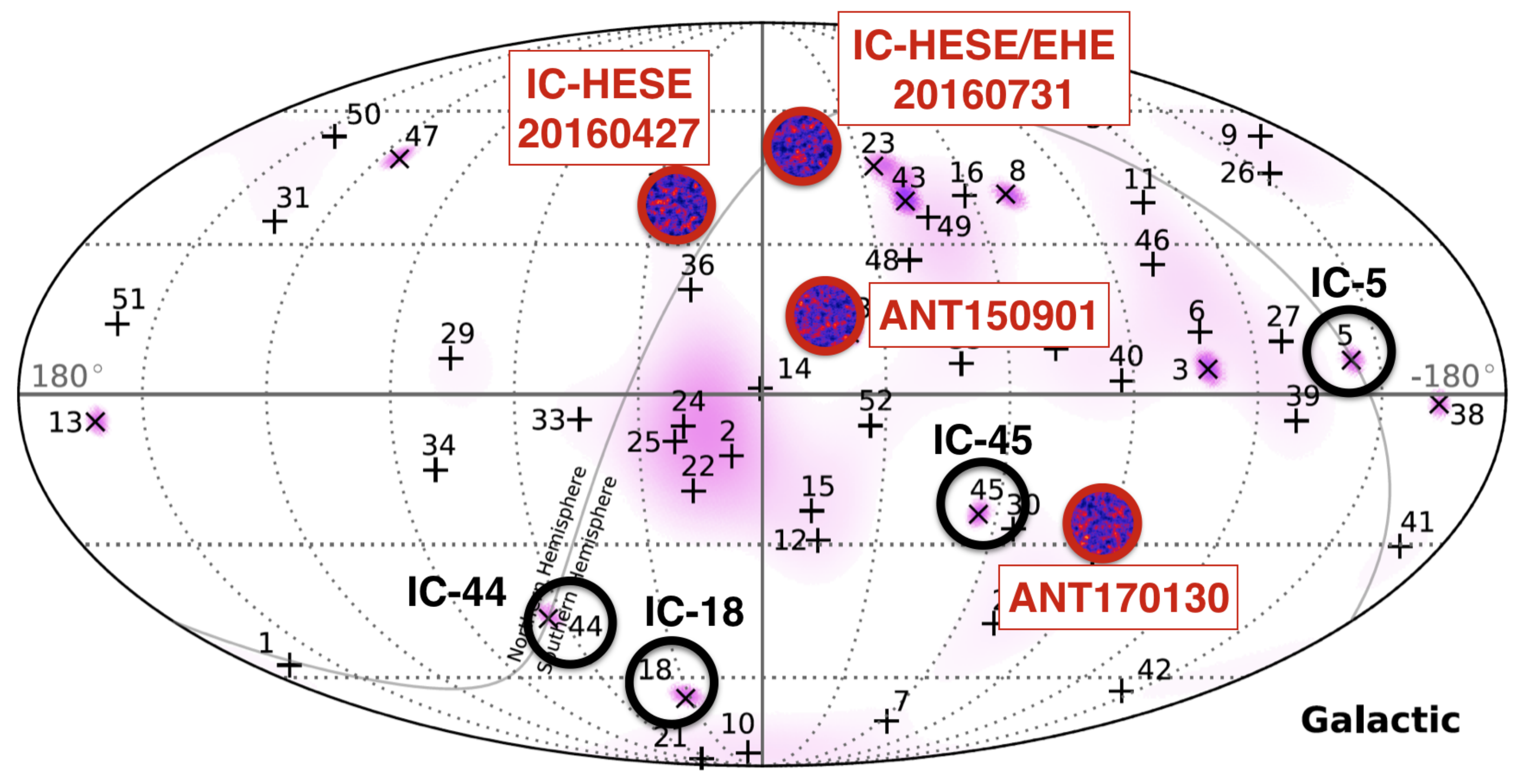}}
\end{minipage}
\caption[]{Left: Schematic view of the H.E.S.S. alert reception system used to react to alerts from observatories covering all wavelengths and astrophysical messengers. From~\cite{HoischenBaikal16, HESS-GRBs-ICRC2017}. Right: The high-energy neutrino sky represented by events detected by IceCube. Archival H.E.S.S. observations around the events IC-5, IC-18, IC-44 and IC-45 are highlighted by black circles. Follow-up observations of neutrino alerts have been performed around the regions denoted in red. Modified from~\cite{IC_HESE_ICRC2015}.}
\label{fig:HESS}
\end{figure}

\subsection{The H.E.S.S. high-energy gamma-ray observatory}
The H.E.S.S. imaging atmospheric Cherenkov telescope array is located at an elevation of 1800 m above sea level on the Khomas Highland plateau of Namibia (23$^{\circ}16'18''$ South, $16^{\circ}30'00''$ East). The original array, inaugurated in 2004, is composed of four telescopes, each with a mirror of 12\,m diameter and a camera holding an array of 960 photomultiplier tubes. 
In 2012 a fifth telescope with a 28\,m diameter mirror was commissioned. 
It was conceived to extend the accessible energy range towards lower energies and to allow for rapid slewing. Its installation was also accompanied by significant efforts to optimize the alert reception and subsequent reaction scheme. The new system is now able to reach every point on the accessible sky and start follow-up observations within minutes~\cite{HESSdrive}. The implemented multi-purpose alert system is connecting the H.E.S.S. observatory to a large variety of observatories worldwide covering all wavelengths and all astrophysical messengers and is thus allowing for an extensive multi-messenger program. A schematic view of the alert system is given in the left plot of Fig.~\ref{fig:HESS}, further details can be found in~\cite{HoischenBaikal16, HESS-GRBs-ICRC2017}.  

\section{High-energy neutrino follow-up observations with H.E.S.S.}


\begin{figure}[!t]
\vspace{-5mm}
\begin{minipage}{0.48\linewidth}
\centerline{\includegraphics[width=0.95\linewidth]{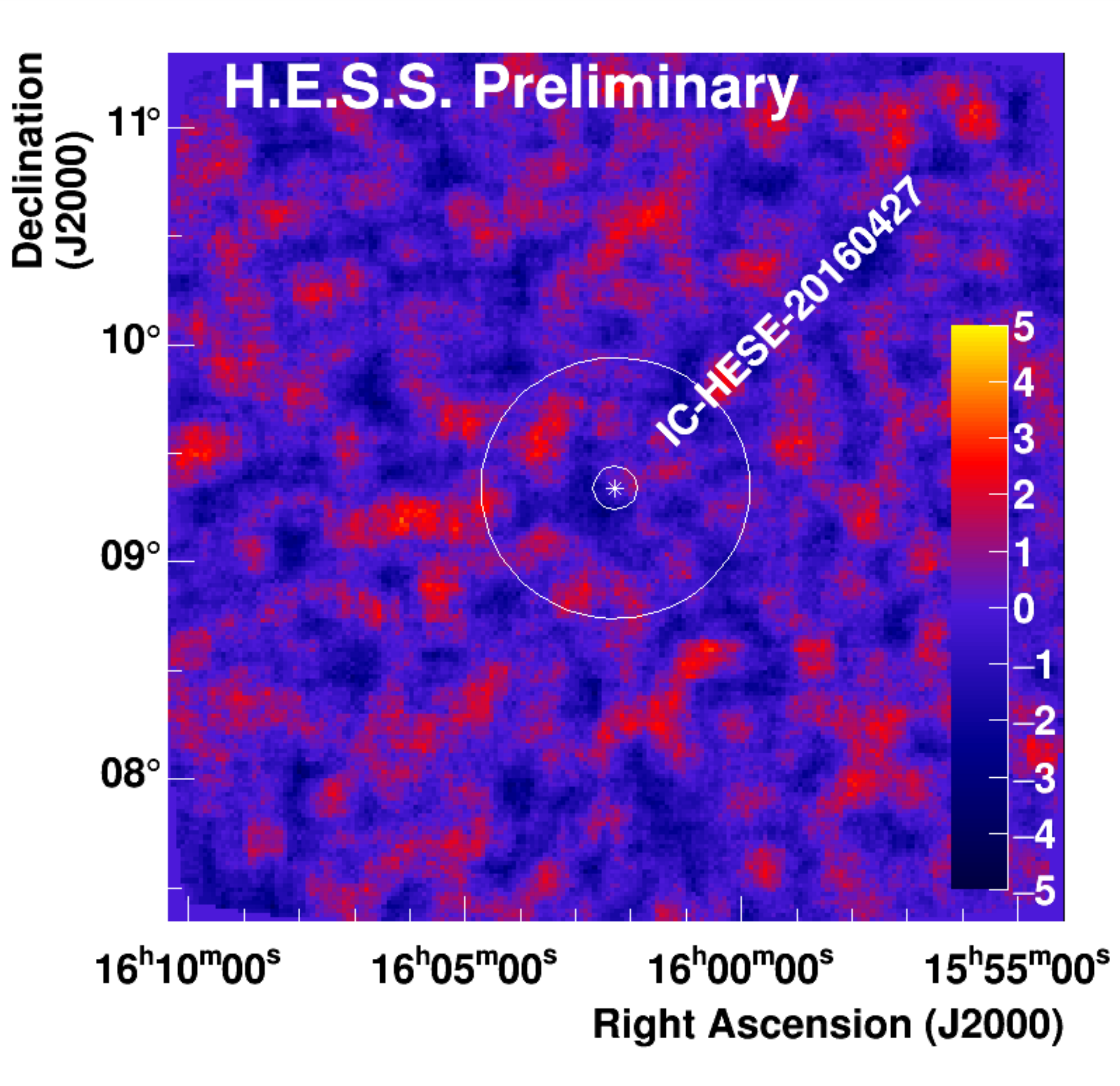}}
\end{minipage}
\hfill
\begin{minipage}{0.48\linewidth}
\centerline{\includegraphics[width=0.95\linewidth]{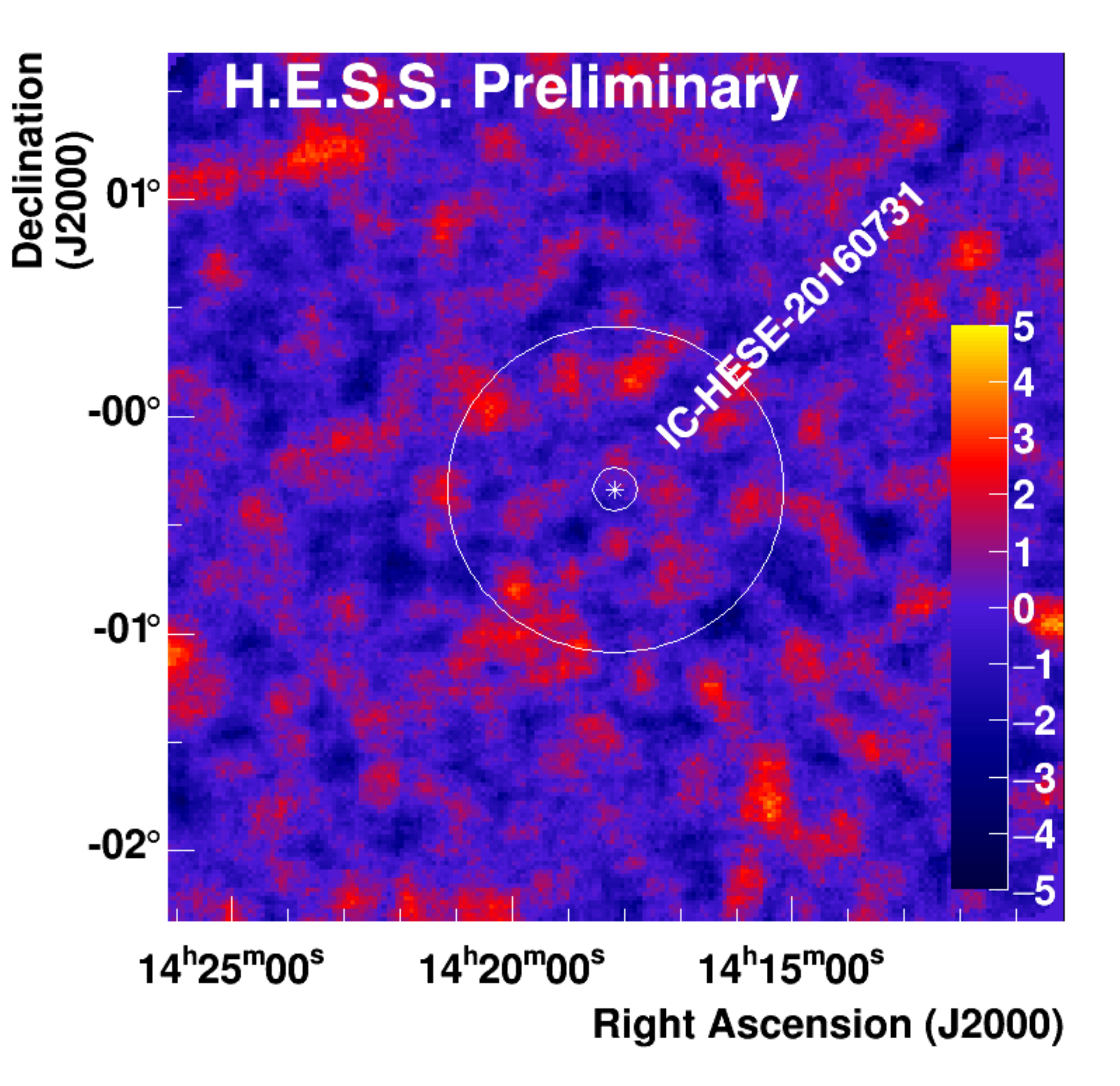}}
\end{minipage}
\caption[]{Significance maps derived from H.E.S.S. observations of the regions around the arrival directions of HESE neutrino alerts emitted by IceCube. The inner circles illustrate the size of the H.E.S.S. PSF and the outer circles denote the individual $90\,\%$ containment angular uncertainty on the neutrino directions.}
\label{fig:SigniMap}
\end{figure}

Two major instruments searching for astrophysical neutrinos are currently in operation: IceCube~\cite{IceCube} at the South Pole and ANTARES~\cite{ANTARES} in the Mediterranean Sea. So far neither of them has found any significant localized excess in the neutrino sky (e.g.~\cite{IC_pointsources}). Yet, a significant breakthrough has been made by the IceCube collaboration through the identification of an astrophysical flux of high-energy neutrinos (see Fig.~\ref{fig:HESS},~\cite{IC_HESE_ICRC2015}). However, the origin of these neutrinos remains unknown. Multi-messenger analyses and especially searches for high-energy gamma-ray counterparts to these events may therefore contribute crucial additional information and may finally help to locate their sources. To allow for timely follow-up observations both neutrino telescopes have implemented systems for rapid event reconstructions, filtering and subsequent alert emission. Whereas the achieved latencies are at the order of several minutes for IceCube~\cite{ICalerts}, the TAToO system of the ANTARES collaboration allows to emit alerts within tens of seconds~\cite{TAToO}. The alert reception system of H.E.S.S. is connected to both neutrino telescopes via direct and dedicated links, allowing for fully automatic exchange of information and subsequent follow-up observations. 

Exploiting this system, we here present first results of searches for coincident gamma-ray emission accompanying high-energy neutrino events. Such a joint detection would point towards transient emission processes involving hadronic particle acceleration and may therefore provide further hints towards the long sought sources of high-energy cosmic rays. If not stated otherwise all data described here have been taken with the full array of all five H.E.S.S. telescopes. They were reconstructed using the Model Analysis~\cite{ModelAnalysis}, an advanced Cherenkov image reconstruction method in which the recorded shower images of all triggered telescopes are compared to a semi-analytical model of gamma-ray showers by means of a log-likelihood optimization. During the analysis we require that data from at least two telescopes participated in the reconstruction of the gamma-ray induced air shower and that the events fulfill a loose cut configuration allowing for comparably low energy thresholds.

\subsection{Searches for steady gamma-ray sources}
Over the last years four H.E.S.S. observation campaigns searched for continuous emission of high-energy gamma rays from regions around neutrino events detected by IceCube, fulfilling the "High-energy Starting Event" (HESE)~\cite{IC_HESE_ICRC2015} criteria. None of the observed regions showed significant high-energy gamma-ray emission. To derive upper limits on the gamma-ray flux while taking into account the large positional uncertainty related to the neutrino events (at the order of $1.2\,\mathrm{deg}$) we produced maps showing integral flux upper limits for the whole region-of-interest (ROI) around the neutrino events. 
Details are given in ~\cite{Multimessenger_MoriondVHEPU2017}. These observations are part of a global effort by all current high-energy gamma-ray observatories. A first joint and combined analysis aiming at deriving constraints on the gamma-ray luminosity and density of the sources responsible for the IceCube flux is presented in~\cite{NeutrinoIACTS_Santander_ICRC2017}.

\begin{figure}[!t]
\vspace{-5mm}
\begin{minipage}{0.48\linewidth}
\centerline{\includegraphics[width=0.95\linewidth]{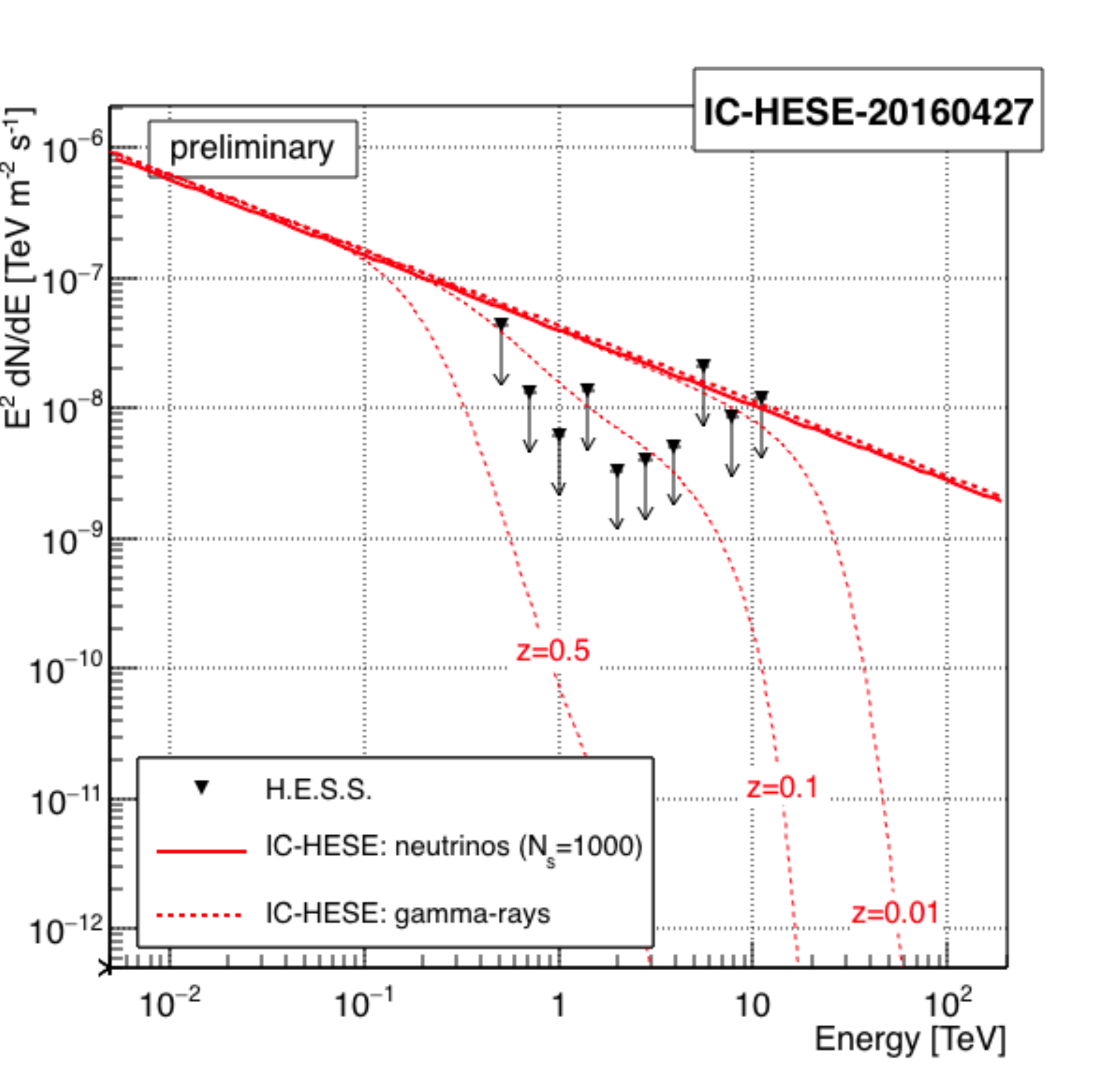}}
\end{minipage}
\hfill
\begin{minipage}{0.48\linewidth}
\centerline{\includegraphics[width=0.95\linewidth]{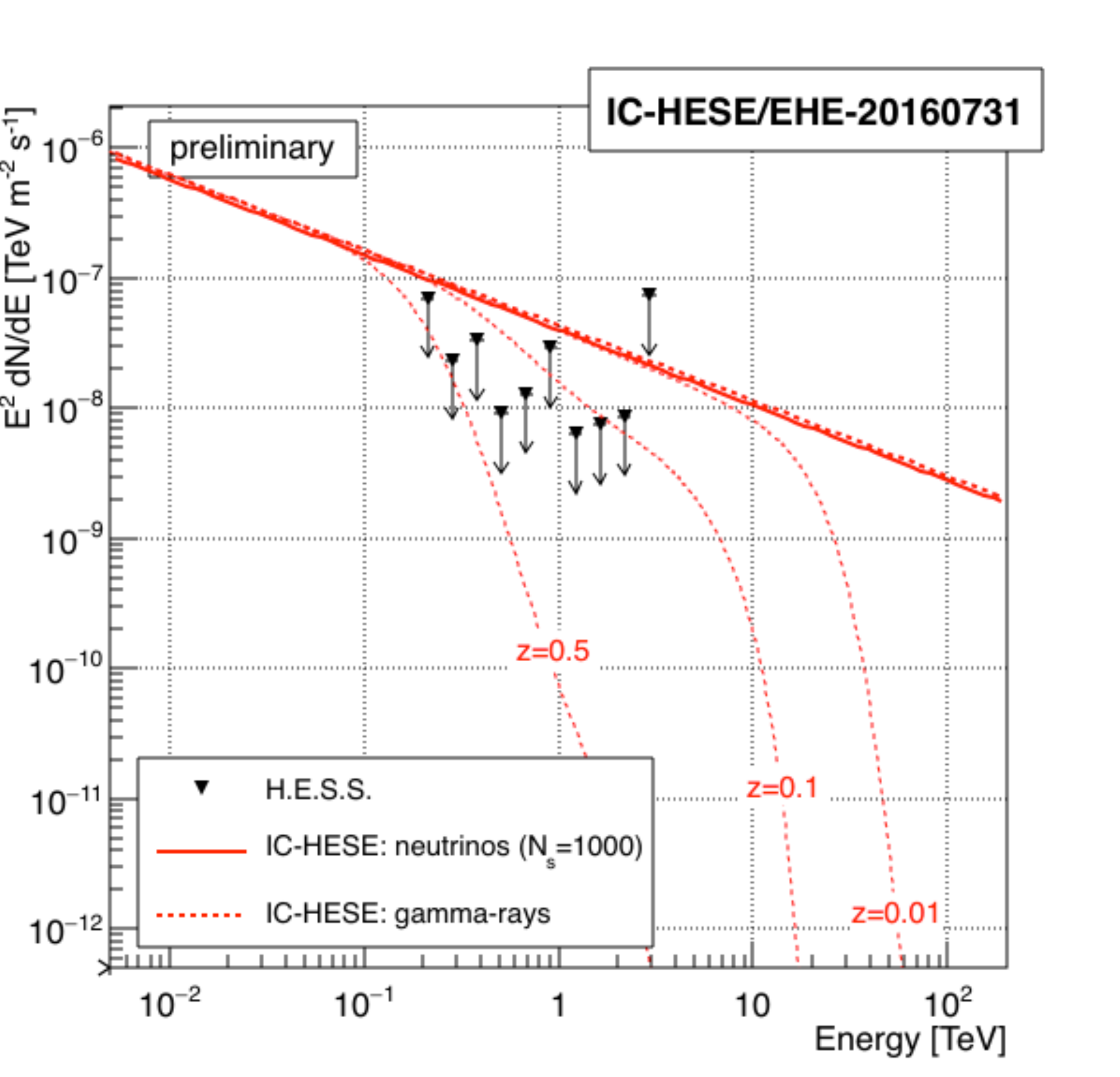}}
\end{minipage}
\caption[]{VHE gamma-ray flux limits $\Phi_\mathrm{UL}$ at 95~\% CL derived from the H.E.S.S. observations (black arrows) assuming a point-like source in the center of neutrino uncertainty region. The estimate of the gamma-ray flux (dashed, red line) has been derived from the IceCube measurement of a diffuse neutrino flux (solid, red line).}
\label{fig:diffLimits}
\end{figure}

\subsection{Searches for transient multi-messenger sources}
The H.E.S.S. multi-messenger program has been recently extended beyond the searches for steady gamma-ray emitters by combining the capabilities of high-energy neutrino telescopes to analyse the data in realtime and emit notifications with the performance of the H.E.S.S. observatory to rapidly react to incoming alerts.

\subsubsection{Follow-up of alerts from IceCube}
On April 27th, 2016, at 05:52:32 UT, IceCube recorded an event fulfilling the HESE criteria and issued an alert soon afterwards. The neutrino has been reconstructed to originate from $\mathrm{RA}=16\mathrm{h}\,02\mathrm{m}\,16\mathrm{s},\; \mathrm{Dec}=9.34\,\mathrm{deg} $ with an uncertainty of $36'$ (radius, $90\%$ containment) and a total charge deposited within the detector of about $18900$ photoelectrons. Visibility and weather constrains delayed H.E.S.S. observations of the identified region to April 29th, 2016 around 21:07 UTC. The observations were taken with an average zenith angle of around $50\,\mathrm{deg}$, the resulting energy threshold, defined as the energy where the acceptance is $10\%$ of its maximum value, is 350\,GeV. Analyzing the total dataset of $1.7\,\mathrm{h}$ effective livetime, the significance map shown in Fig.~\ref{fig:SigniMap} has been obtained. As no significant gamma-ray emission could be found, differential upper limits on the gamma-ray flux from the central position of the covered ROI (cf. the inner circles in Fig.~\ref{fig:SigniMap}) have been calculated. They are shown as black arrows in Fig.~\ref{fig:diffLimits}. For illustration, the obtained limits  are compared with expectations derived from the IceCube diffuse, all-sky neutrino flux ($\mathrm{E}^2 \times \Phi(\mathrm{E}) = 2.2 \pm 0.7 \times 10^{-8} (\mathrm{E}/100\,\mathrm{TeV})^{-0.58}\,\mathrm{GeV}\,\mathrm{cm}^{-2}\,\mathrm{s}^{-ˆ'1}\,\mathrm{sr}^{-ˆ'1}$,~\cite{IC_HESE_ICRC2015}) denoted by the solid red lines in Fig.~\ref{fig:diffLimits}. The all-sky flux was distributed over 1000 putative sources, a number currently not excluded by the searches for point-like neutrino sources. The derived flux can therefore be considered as an upper-bound on the expected neutrino flux per contributing source. The conversion into a gamma-ray flux (dashed red lines in Fig.~\ref{fig:diffLimits}) is relying on the parametrization of $pp$ interactions within or close to a generic hadronic accelerator as given in~\cite{Kappes_NuGammaFlux}.

\begin{figure}[!t]
\vspace{-5mm}
\begin{minipage}{0.48\linewidth}
\centerline{\includegraphics[width=1.0\linewidth]{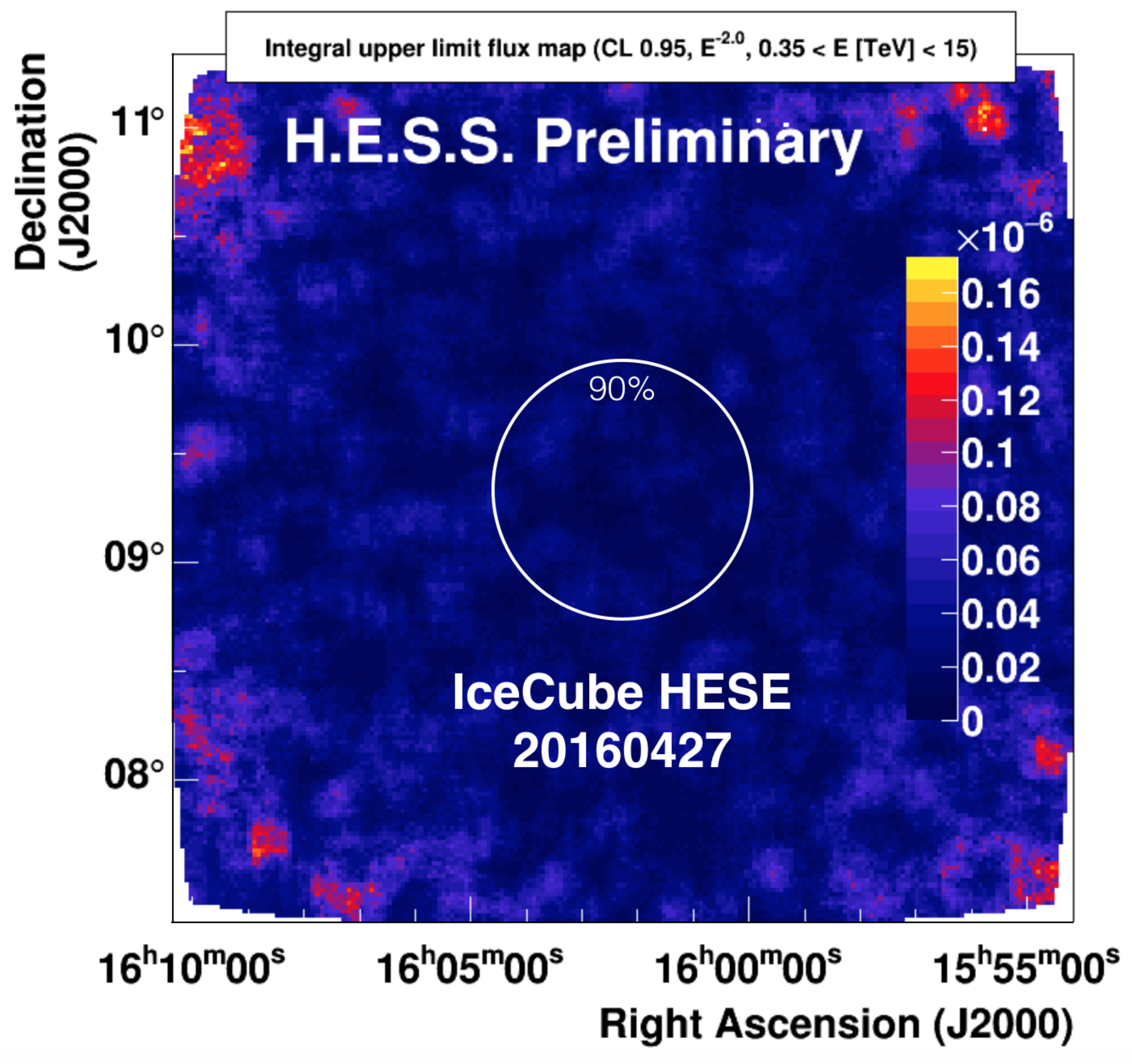}}
\end{minipage}
\hfill
\begin{minipage}{0.48\linewidth}
\centerline{\includegraphics[width=0.95\linewidth]{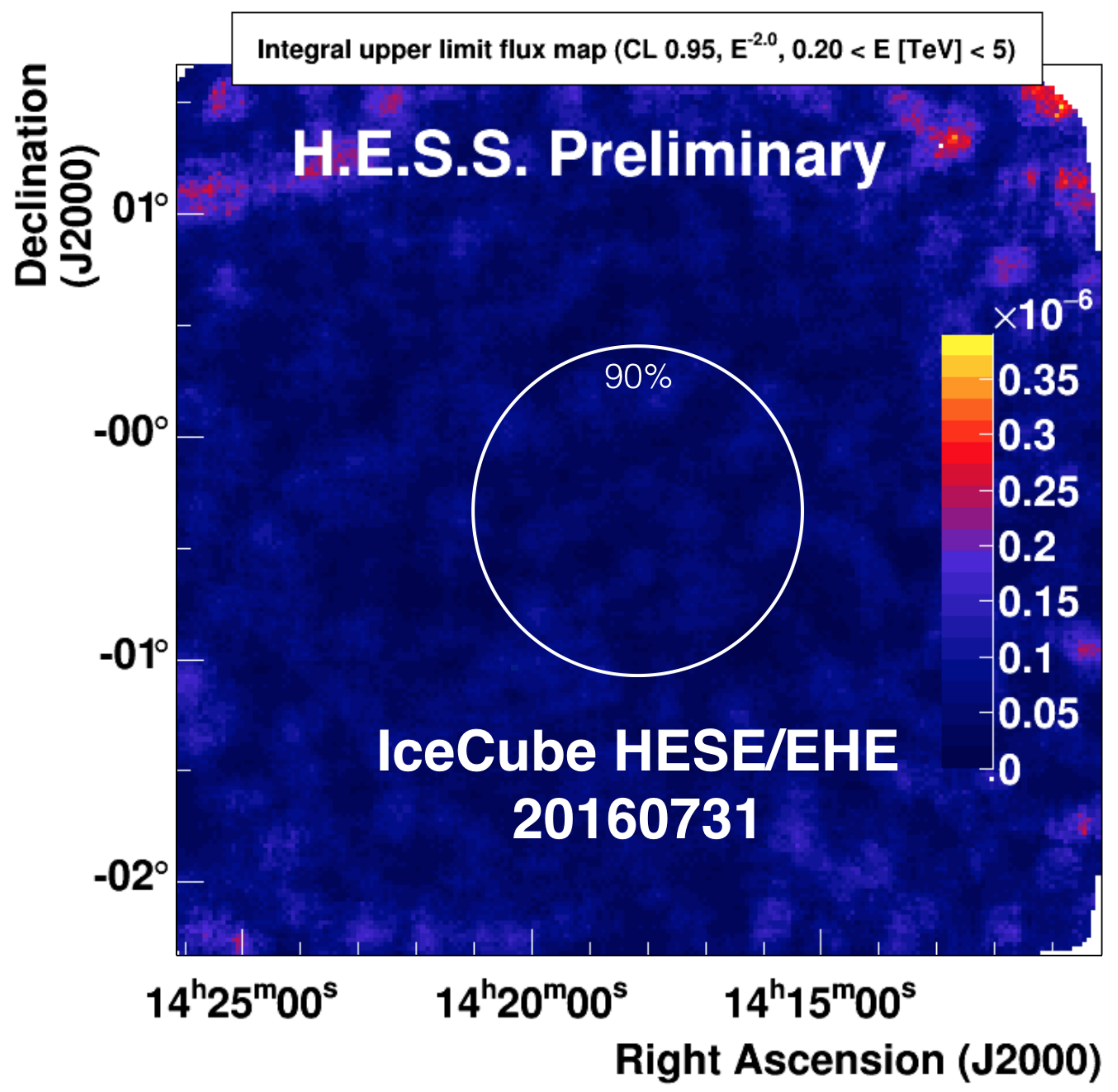}}
\end{minipage}
\caption[]{Integral VHE gamma-ray flux limits ($95\,\%$ C.L., assuming a $\mathrm{E}^{-2}$ spectrum) in units of $\mathrm{m}^{-2} \mathrm{s}^{-1}$ derived from H.E.S.S. follow-up observations of IceCube neutrino alerts. The white circles denote the angular uncertainty ($90\,\%$ containment ) on the neutrino directions.}
\label{fig:ULmap}
\end{figure}

A second observation campaign started after the detection of a neutrino event fulfilling both the HESE and the EHE criteria recorded by IceCube on July 31st, 2016 at 01:55:04 UT with a total deposited charge of about $15800$ photoelectrons. H.E.S.S. observations started the same day around 17:50 UTC as soon as the reconstructed neutrino direction ($\mathrm{RA}=14\mathrm{h}\,18\mathrm{m}\,11\mathrm{s},\; \mathrm{Dec}=-0.33\,\mathrm{deg} $ with an uncertainty of $45'$ radius at $90\%$ containment) became visible from the H.E.S.S. site. 
As can be seen in the significance map shown in the right plot of Fig.~\ref{fig:SigniMap}, no significant gamma-ray emission has been found. 
Following the procedure outlined above, differential upper limits on the gamma-ray flux for the central position of the ROI (cf. Fig.~\ref{fig:diffLimits}) and integral upper limits on the full region (cf. Fig~\ref{fig:ULmap}) have been derived.

\subsubsection{Follow-up of alerts from ANTARES}
\begin{figure}[!t]
\vspace{-5mm}
\begin{minipage}{0.54\linewidth}
\centerline{\includegraphics[width=1.0\linewidth]{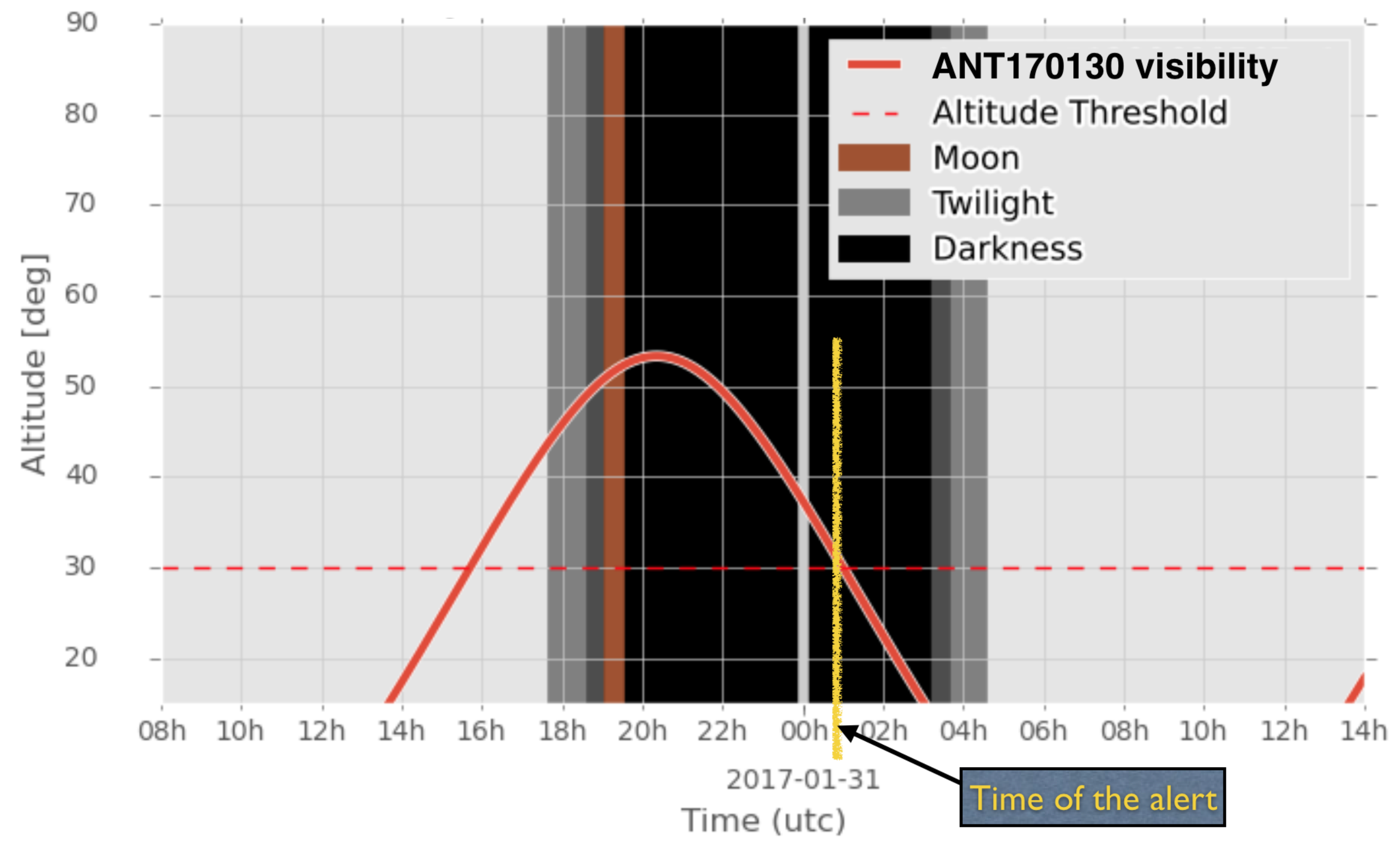}}
\end{minipage}
\hfill
\begin{minipage}{0.45\linewidth}
\centerline{\includegraphics[width=0.95\linewidth]{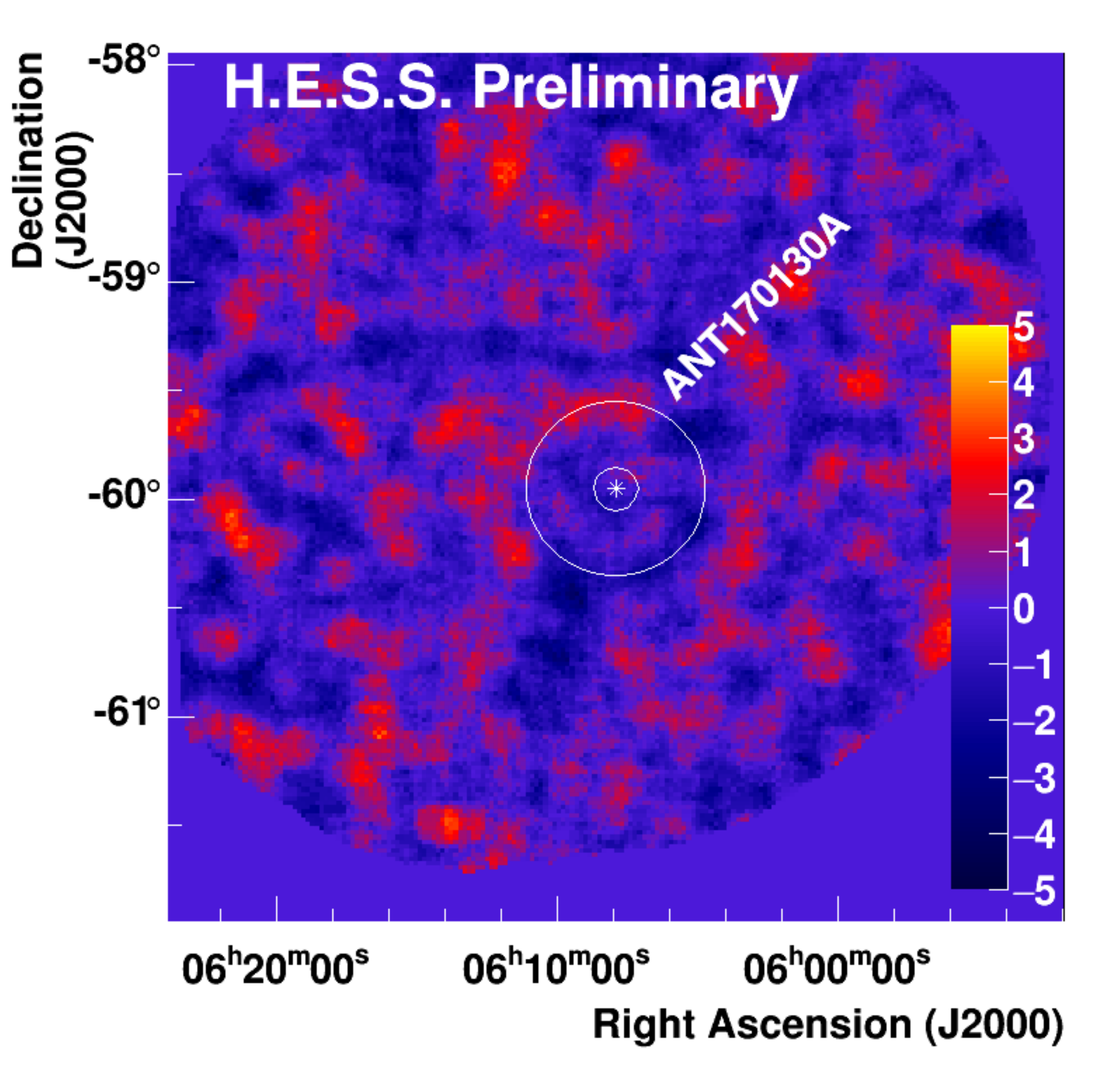}}
\end{minipage}
\caption[]{Left: Visibility of the ANTARES alert received 2017-01-30 at the H.E.S.S. site. Right: Significance map derived from H.E.S.S. observations of the regions around the ANTARES alert ANT170130A. The inner circle illustrates the size of the H.E.S.S. PSF and the outer circle denotes the $50\,\%$ containment angular uncertainty on the neutrino direction of $0.4\,\mathrm{deg}$. }
\label{fig:ANT170130}
\end{figure}

A dedicated direct link between the ANTARES online reconstruction and alert emission system TAToO~\cite{TAToO} and the H.E.S.S. alert system allows for rapid exchanges of alerts between the two experiments. The performance of this approach has been illustrated by an alert received on January 30, 2017. On 00:39:12 ANTARES recorded a single, high-energy neutrino. The data transfer to the shore station, their filtering, the event reconstruction triggered a VoEvent notice to be emitted to partner observatories at 00:39:25. The alert was received at the H.E.S.S. site triggering automatic follow-up observations which were started at 00:39:44, i.e. only 32 seconds after the neutrino event. About one hour of H.E.S.S. observations could be obtained immediately following the alert before the direction moved outside the visibility window (cf. Fig.~\ref{fig:ANT170130}). Searches for transient gamma-ray emission on various timescales is in progress. Another 30min of data was taken during the night after the alert (2017-01-31 19:29 UT) in order to search for afterglow and/or delayed or emission. A preliminary analysis using monoscopic data from the 28m telescope and time integrating the full dataset did not reveal any significant gamma-ray emission. The significance map of the region around the reconstructed neutrino direction is shown in Fig.~\ref{fig:ANT170130}.

\section{Gravitational wave follow-up observations with H.E.S.S.}

After substantial upgrades and improvements, the Advanced Ligo interferometers detected the first gravitational waves (GWs) end of 2015~\cite{LigoDetectionPaper}. This breakthrough has opened a new window and completed the multi-messenger picture of particles and radiation available to access the high energy universe. The localisation of the massive binary black hole systems whose mergers caused the emission of the observed GWs was unfortunately rather imprecise. Nevertheless, a large number of observatories covering all wavelengths from radio to gamma rays are participating in a global effort to search for electromagnetic (and high-energy neutrino) counterparts to the GW events. The H.E.S.S. experiment is participating in this effort. To cope with the large localisation uncertainties (hundreds to thousands of square degrees) we developed dedicated follow-up strategies, implemented within the automatic alert reception system mentioned above. The general approach is outlined here, details are given in~\cite{SeglarArroyo_Moriond2017}. Results from first H.E.S.S. searches for high-energy gamma-ray counterparts to GWs will be reported elsewhere.

Our approach to determine the optimal pointing strategy following the detection of a GW uses the full three dimensional information on the localisation of the event provided by the Virgo/Ligo interferometers~\cite{GoingTheDistance} combined with a catalog of galaxies (GLADE~\cite{GLADE}). Several algorithms have been developed to select the most promising pointing directions from the resulting 3D probability distribution. All of them take into account the time-dependent visibility from the site of the H.E.S.S. observatory and optimize the zenith angle of the observations in order to achieve a low energy threshold. 

A first algorithm is optimized for online reactions to incoming alerts and produces the scheduling for a full night within a few minutes. This {\it one-galaxy} approach starts by ordering all galaxies in the catalog (passing an adjustable threshold) following their likelihood to be associated with the GW event and picks the direction of the most likely one as first target. Thanks to the large field-of-view of the H.E.S.S. telescope system of about $2.5\,\mathrm{deg}$ radius, several of the highly likely galaxies will be covered by each observation and are removed from the list before selecting the next pointing in an iterative process. 

A second algorithm has been optimized in order to achieve the best coverage of the provided GW uncertainty region and is used for less time constrained scheduling (e.g. reception of the alert outside the H.E.S.S. observation time, update of the scheduling for a second or third night of follow-up observations). While relying on the same 3D {\it galaxy $\times$ GW} probability map, this {\it galaxies-in-FoV} approach takes into account all galaxies within the field-of-view in order to assign a probability coverage to each potential pointing direction. After choosing the highest priority pointing for the first observation, this probability is re-assessed considering potential overlapping observations, the time dependent evolution of the zenith angle contrains, etc. 

\begin{figure}[t!]
\vspace{-5mm}
\centerline{\includegraphics[width=1.0\linewidth]{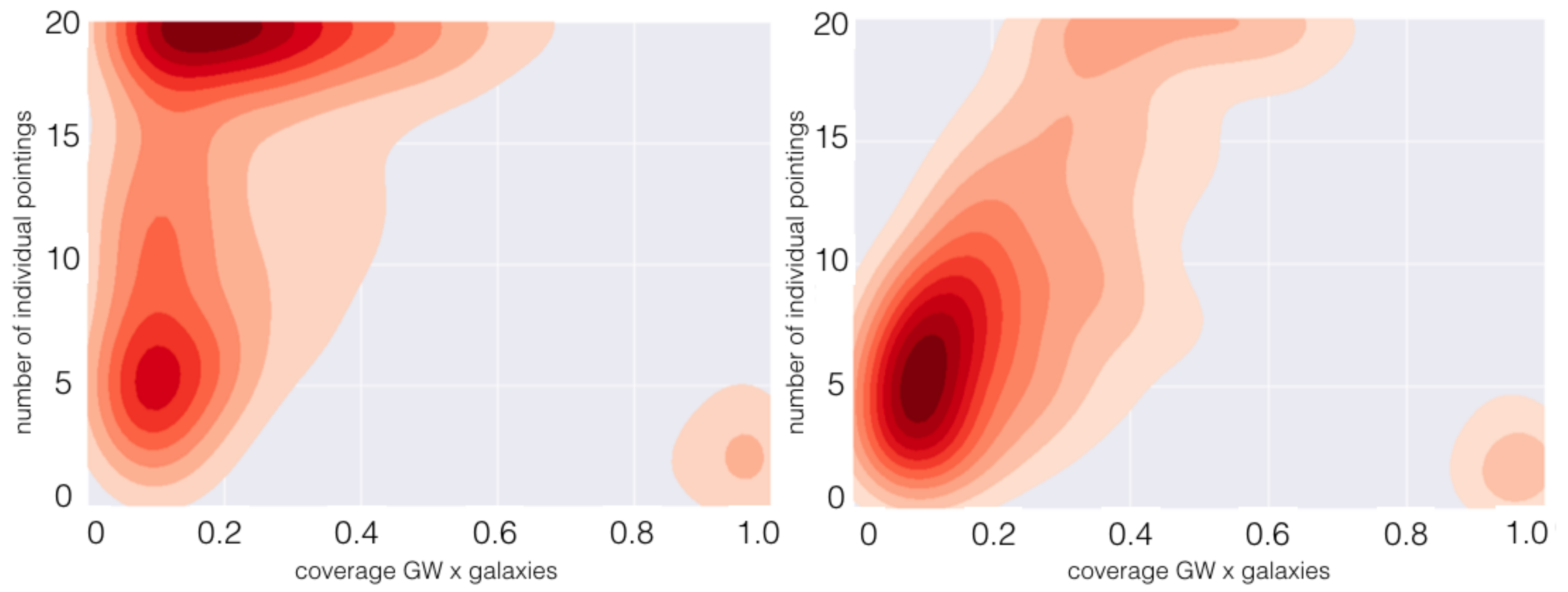}}
\caption[]{Performance of the H.E.S.S. scheduling algorithms developed for follow-up observations of gravitational waves derived from simulated GW events reproducing the O2 run of Adv.Ligo/Adv.Virgo. The rapid {\it one-galaxy} approach (left plot) achieves significant coverages of the GW uncertainty regions whereas the {\it galaxies-in-FoV} algorithm yields similar results already with very few individual pointings. From~\cite{SeglarArroyo_Moriond2017}. }
\label{fig:GWperformance}
\end{figure}

The performance of the scheduling algorithms has been tested by injecting a set of simulated neutron star binary mergers. The response of Adv. Ligo and Adv. Virgo has been simulated following the expected performance during their second observation run O2. Allowing for up to 20 individual pointings performed over a maximum of three consecutive nights we achieve an average coverage of the combined {\it galaxy $\times$ GW} probability of more than $20\,\%$. As illustrated in Fig.~\ref{fig:GWperformance}, the more time consuming {\it galaxies-in-FoV} algorithm reaches this goal significantly faster. A combination of both algorithms, also including an additional prior favoring galaxies with high blue luminosity as indicator of a high star formation rate, is currently under study. 

\section{Acknowledgments}
{\footnotesize The support of the Namibian authorities and of the University of Namibia in facilitating the construction and operation of H.E.S.S. is gratefully acknowledged, as is the support by the German Ministry for Education and Research (BMBF), the Max Planck Society, the German Research Foundation (DFG), the French Ministry for Research, the CNRS-IN2P3 and the Astroparticle Interdisciplinary Programme of the CNRS, the U.K. Science and Technology Facilities Council (STFC), the IPNP of the Charles University, the Czech Science Foundation, the Polish Ministry of Science and Higher Education, the South African Department of Science and Technology and National Research Foundation, the University of Namibia, the Innsbruck University, the Austrian Science Fund (FWF), and the Austrian Federal Ministry for Science, Research and Economy, and by the University of Adelaide and the Australian Research Council. We appreciate the excellent work of the technical support staff in Berlin, Durham, Hamburg, Heidelberg, Palaiseau, Paris, Saclay, and in Namibia in the construction and operation of the equipment. This work benefited from services provided by the H.E.S.S. Virtual Organisation, supported by the national resource providers of the EGI Federation.\\
We acknowledge valuable discussions facilitating the preparation of the H.E.S.S. observations discussed here with colleagues from the IceCube Collaboration: E. Bernardini, J. Dumm, W. Giang, A. Stasik, T. Kintscher and C. Kopper as well as M. Ageron, D. Dornic and A. Mathieu from the ANTARES Collaboration.\\


\begin{thebibliography}{99}
{\footnotesize 

\bibitem{HESSdrive} P. Hofverberg et al. (H.E.S.S. Collaboration), 33rd ICRC (2013), \arxiv{1307.4550}.\vspace{-2mm}

\bibitem{HoischenBaikal16} C. Hoischen et al. (H.E.S.S. Collaboration), Baikal Three Messenger Conference (2016).\vspace{-2mm}

\bibitem{HESS-GRBs-ICRC2017} C. Hoischen et al. (H.E.S.S. Collaboration), \pos{PoS(ICRC2017)636}\vspace{-2mm}

\bibitem{IceCube} R. Abbasi et al. (IceCube Collaboration), NIM A 601 (2009) 294.\vspace{-2mm}

\bibitem{ANTARES} M. Ageron et al. (ANTARES Collaboration), NIM A 656 (2011) 11.\vspace{-2mm}

\bibitem{IC_pointsources} M. G. Aartsen et al. (IceCube collaboration), APJ 835 (2017) 151.\vspace{-2mm}

\bibitem{IC_HESE_ICRC2015} C. Kopper et al. (IceCube Collaboration),  \pos{PoS(ICRC2015)1081}.\vspace{-2mm}

\bibitem{ICalerts} M.G. Aartsen et al. (IceCube Collaboration), APP 92 (2017) 30-41.\vspace{-2mm}

\bibitem{TAToO} M. Ageron et al. (ANTARES Collaboration), Astropart. Phys. 35 (2012) 530. \vspace{-2mm}

\bibitem{ModelAnalysis} M. de Naurois, M., L. Rolland, Astropart. Physics 32 (2009) 231.\vspace{-2mm}

\bibitem{Multimessenger_MoriondVHEPU2017}  F. Sch\"ussler et al. (H.E.S.S. Collaboration), Rencontres de Moriond VHEPU (2017), \arxiv{1705.08258}. \vspace{-2mm}

\bibitem{NeutrinoIACTS_Santander_ICRC2017} M. Santander, D. Dorner, J. Dumm, K. Satalecka and F. Sch\"ussler (VERITAS, FACT, IceCube, MAGIC and H.E.S.S. Collaborations), \pos{PoS(ICRC2017)618}\vspace{-2mm} 

\bibitem{Kappes_NuGammaFlux} A. Kappes, J. Hinton, C. Stegmann and F. Aharonian, APJ 656, 870?878 (2007).\vspace{-2mm}

\bibitem{LigoDetectionPaper} B.P. Abbott et al. (LIGO and Virgo Collaborations), Phys. Rev. Lett. 116 (2016) 241103.\vspace{-2mm}

\bibitem{SkymapViewer} R. Williams, T. Boch and K. Li, {\it Skymap viewer}, {\footnotesize \url{https://losc.ligo.org/s/skymapViewer/}}\vspace{-2mm}

\bibitem{SeglarArroyo_Moriond2017}  M. Seglar-Arroyo and F. Sch\"ussler (H.E.S.S. Collaboration), Rencontres de Moriond VHEPU (2017), \arxiv{1705.10138}.\vspace{-2mm}

\bibitem{GoingTheDistance} Singer, LP. et al., APJ Letters 829.1 (2016): L15, \arxiv{1603.0733}.\vspace{-2mm}

\bibitem{GLADE} G. Dalya et al., GLADE catalog, {\footnotesize \url{http://aquarius.elte.hu/glade}}\vspace{-2mm}
}
\end{thebibliography}
\end{document}